\documentclass[aps,prd,twocolumn,showpacs,nofootinbib]{revtex4}
\usepackage[dvips]{graphicx} \usepackage{amssymb} \usepackage{psfrag}
\begin{document}
\title{$U(1)_{B-L}$: Neutrino Physics and Inflation}
\thanks{Based on talks given by Q. Shafi at the 6th European Meeting From The Planck Scale To The Electroweak Scale  
  (Planck 03), Madrid, Spain, May 26-31, 2003; at the 11th International Workshop on Neutrino Telescopes, Venice, Italy, 22-25 Feb 2005; and at the 11th International Symposium on Particles, Strings and Cosmology (PASCOS 2005), Gyeongju, Korea, 30 May - 4 Jun 2005.}
\author{V. N. {\c S}eno$\breve{\textrm{g}}$uz} \email{nefer@udel.edu} \author{Q. Shafi} \email{shafi@bxclu.bartol.udel.edu}  
\affiliation{Bartol Research Institute, Department of Physics and Astronomy, University of Delaware, Newark, DE
19716, USA} 
\begin{abstract} 
 A gauged $U(1)_{B-L}$ symmetry predicts three right handed handed neutrinos and its
spontaneous breaking automatically yields the seesaw mechanism. In a
supersymmetric setting this breaking can be nicely linked with inflation to
yield $\delta T/ T$ proportional to $(M_{B-L} / M_P)^2$, where $M_{B-L}$ $(M_P)$ denote the
$B-L$ breaking (Planck) scale. Thus $M_{B-L}$ is estimated to be of
order $10^{16}$ GeV, and the heaviest right handed neutrino mass is less than or of order $10^{14}$
GeV. A second right handed neutrino turns out to  have a mass of order $10-10^2\
T_r$, where $T_r$ ($\lesssim10^{9}$ GeV)  denotes the reheat temperature.  A $U(1)$ R
symmetry plays an essential role in implementing inflation and leptogenesis,
resolving the MSSM $\mu$ problem and eliminating dimension five nucleon decay. An
unbroken $Z_2$ subgroup plays the role of matter parity. The scalar spectral index $n_s=0.99\pm0.01$
for the simplest models, while in smooth hybrid inflation $n_s\ge0.97$.
The tensor to scalar ratio $r$ is negligible, and $\textrm {d}n_{s}/\textrm {d}\ln k\lesssim10^{-3}$.
\end{abstract}
\maketitle

\section{Introduction}
Physics beyond the Standard Model (SM) is required by the following experimental
observations:

\begin{itemize}
\item Neutrino Oscillations: $\Delta m_{\rm SM}^2 \lesssim 10^{-10}\ {\rm eV}^2\ll$ (mass difference)$^2$
                          needed to understand atmospheric and solar neutrino
                          observations;
\item CMB Anisotropy ($\delta T/T$): requires inflation which cannot be realized in the
                             SM;
\item Non-Baryonic Dark Matter ($\Omega_{\rm CDM} = 0.25$): SM has no plausible candidate;
\item Baryon Asymmetry ($n_b/s\sim10^{-10}$): Not possible to achieve in the SM.
\end{itemize}

Recall that at the renormalizable level the SM possesses a global $U(1)_{B-L}$
symmetry. If the symmetry is gauged, the anomaly cancellation requires the
existence of three right handed neutrinos. An important question therefore
is the symmetry breaking scale of $U(1)_{B-L}$. Note that this scale is not fixed
by the evolution of the three SM gauge couplings. Remarkably, we will be able
to determine the $M_{B-L}$ by implementing inflation. With $M_{B-L}$ well below the
Planck scale the seesaw mechanism enables us to realize light neutrino masses
in the desired range. Furthermore, it will turn out that leptogenesis is a
natural outcome after inflation is over.

The introduction of a gauge $U(1)_{B-L}$ symmetry broken at a scale well below the
Planck scale exacerbates the well known gauge hierarchy problem. There are at
least four potential hierarchy problems one could consider:
\begin{itemize}
\item $M_W \ll M_P$;
\item $M_{B-L} \ll M_P$ (required by neutrino oscillations);
\item $m_{\chi} \ll M_P$ (where $m_{\chi}$ denotes the inflaton mass);
\item $f_a \sim 10^{10}-10^{12}\ {\rm GeV}\ (\ll M_P)$, where $f_a$ denotes the axion decay constant.
\end{itemize}
Supersymmetry (SUSY) can certainly help here, especially if the SUSY breaking
scale in the observable sector is of order TeV. Thus, it seems that a good
starting point, instead of SM$\,\times U(1)_{B-L}$, could be MSSM$\,\times U(1)_{B-L}$.
The $Z_2$ `matter' parity associated with the MSSM has two important consequences. It
eliminates rapid (dimension four) proton decay, and it delivers a respectable
cold dark matter candidate in the form of LSP. However, Planck scale suppressed
dimension five proton decay is still present and one simple solution is to
embed $Z_2$ in a $U(1)_R$ symmetry. It turns out that the R symmetry also plays an
essential role in realizing a compelling inflationary scenario and in the
resolution of the MSSM $\mu$ problem. Finally it seems natural to extend the above
discussion to larger groups, especially to $SO(10)$ and its various subgroups.

\section{Supersymmetric Hybrid Inflation Models}
\label{chap2}
In this section, we review a class of supersymmetric hybrid inflation models  
\cite{Senoguz:2003zw,Senoguz:2004vu,Lazarides:2001zd}
where inflation can be linked to the breaking of $U(1)_{B-L}$.
We compute the allowed range of the
dimensionless coupling in the superpotential and the dependence of the
spectral index on this coupling, in the presence of canonical supergravity (SUGRA)
corrections.

The simplest
supersymmetric hybrid inflation model \cite{Dvali:1994ms} is realized by the
renormalizable superpotential \cite{Copeland:1994vg}
\begin{equation} \label{super} W_1=\kappa S(\Phi\overline{\Phi}-M^{2})
\end{equation}
\noindent where $\Phi(\overline{\Phi})$ denote a conjugate pair of superfields
transforming as nontrivial representations of some gauge group $G$, $S$ is a
gauge singlet superfield, and $\kappa$ $(>0)$ is a dimensionless coupling.  A
suitable $U(1)$ R-symmetry, under which $W_1$ and $S$ transform the same way,
ensures the uniqueness of this superpotential at the renormalizable level \cite{Dvali:1994ms}.
In the absence of supersymmetry breaking, the potential
energy minimum corresponds to non-zero vacuum expectation values (VEVs) $(=M)$ in the
scalar right handed neutrino components
$\big|\langle\nu^c_H\rangle\big|=\big|\langle\overline{\nu}^c_H\rangle\big|$ for
$\Phi$ and $\overline{\Phi}$, while the VEV of $S$ is
zero.  (We use the same notation for superfields and their scalar components.)
Thus, $G$ is broken to some subgroup $H$ which, in many interesting models,
coincides with the MSSM gauge group.

In order to realize inflation, the scalar fields $\Phi$,
$\overline{\Phi}$, $S$ must be displayed from their present minima.  For
$|S|>M$, the $\Phi$, $\overline{\Phi}$ VEVs both vanish so that the gauge
symmetry is restored, and the tree level potential energy density
$\kappa^{2}M^{4}$ dominates the universe, as in the originally
proposed hybrid inflation scenario \cite{Linde:1991km,Liddle:1993fq,Copeland:1994vg}.
With supersymmetry thus broken, there are radiative corrections from the
$\Phi-\overline{\Phi}$ supermultiplets that provide logarithmic corrections
to the potential which drives inflation.

In one loop approximation the inflationary effective potential is given
by \cite{Dvali:1994ms}
{\setlength\arraycolsep{2pt}
\begin{eqnarray} \label{loop} V_{\mathrm{LOOP}}=\kappa^{2}M^{4}\bigg[1
+\frac{\kappa^{2}\mathcal{N}}{32\pi^{2}} \Big(
2\ln\frac{\kappa^{2}|S|^{2}}{\Lambda^{2}}+\nonumber\\(z+1)^{2}\ln(1+z^{-1}) 
+(z-1)^{2}\ln(1-z^{-1})\Big) \bigg]\,, \end{eqnarray}}
\noindent where $z\equiv x^{2}\equiv|S|^{2}/M^{2}$, $\mathcal{N}$ is the dimensionality of
the $\Phi$, $\overline{\Phi}$ representations, and $\Lambda$ is
a renormalization mass scale.

The scalar spectral index $n_s$ is given by \cite{Liddle:1992wi,Liddle:1993fq}
\begin{equation} n_s\cong1-6\epsilon+2\eta,\qquad\epsilon\equiv
\frac{m^2_P}{2}\left(\frac{V'}{V}\right)^2,\qquad\eta\equiv\frac{m^2_P V''}{V},
\end{equation}
\noindent where $m_P\simeq2.4\times10^{18}$ GeV is the reduced Planck mass ($M_P/\sqrt{8\pi}$).
We are going to use units $m_P=1$, although sometimes we will
write $m_P$ explicitly.
The primes denote derivatives with respect to the normalized
real scalar field $\sigma\equiv\sqrt{2}|S|$. For relevant values of the
parameters ($\kappa\ll1$), the slow roll conditions ($\epsilon$, $\eta\ll1$) are violated only
`infinitesimally'
close to the critical point at $x=1$ ($|S|=M$) \cite{Lazarides:2001zd}.
So inflation continues practically until this point is
reached, where it abruptly ends.

The number of $e$-folds after the comoving scale $l$ has crossed the horizon is
given by
\begin{equation} \label{efold1}
N_l=\frac{1}{m^2_P}\int^{\sigma_l}_{\sigma_f}\frac{V\rm{d}\sigma}{V'} \end{equation}
where $\sigma_l$ is the value of the field at the comoving scale $l$ and $\sigma_f$
is the value of the field at the end of inflation. Using Eqs. (\ref{loop}, \ref{efold1}),
we obtain
\begin{equation} \label{m_kap}
\kappa\approx\frac{2\sqrt{2}\pi}{\sqrt{\mathcal{N}N_{0}}}\,y_{0}\,\frac{M}{m_{P}}\,.
\end{equation}
Here, the subscript 0 implies that the values correspond to the comoving
scale $l_0=2\pi/k_0$, with $k_0\equiv0.002$ Mpc$^{-1}$.
$N_0\approx55$ is the number of $e$-folds\footnote{$N_0\simeq54+(1/3)\ln(T_r/10^{9}\
 \rm{GeV})$+$(2/3)\ln(v/10^{14}\ \rm{GeV})$, where $T_r$ is the reheat temperature and $v^4$ is
the false vacuum energy density.} and
\begin{equation} \label{yq} y_{0}^{2}=\int_{1}^{x^{2}_{0}}\frac{\textrm{d} z}{z
f(z)}\quad ,y_0\ge 0\,,  \end{equation}
with
\begin{equation} \label{feza}
f(z)=\left(z+1\right)\ln\left(1+z^{-1}\right)+\left(z-1\right)\ln\left(1-z^{-1}\right)\,.
\end{equation}
The amplitude of the
curvature perturbation $\mathcal{P^{{\rm 1/2}}_R}$ is given by\footnote{Note that the quadrupole CMB
anisotropy $\delta T/ T=\mathcal{P^{{\rm 1/2}}_R}/2\sqrt{15}$.}
\begin{equation} \label{perturb}
\mathcal{P^{{\rm 1/2}}_R}=\frac{1}{2\sqrt{3}\pi m^3_P}\frac{V^{3/2}}{|V'|}\,.
\end{equation}
Using Eqs. (\ref{loop}, \ref{m_kap}, \ref{perturb}),
$\mathcal{P^{{\rm 1/2}}_R}$ is found to be \cite{Dvali:1994ms,Lazarides:2001zd,Lazarides:1997dv}
\begin{equation} \label{quad}
\mathcal{P^{{\rm 1/2}}_R}\approx2\left(\frac{N_0}{3\mathcal{N}}\right)^{1/2}
\left(\frac{M}{m_{P}}\right)^{2}x_{0}^{-1}y_{0}^{-1}f (x^{2}_{0})^{-1}\,.
\end{equation}

Up to now, we ignored supergravity (SUGRA) corrections to the potential.
More often than not, SUGRA corrections tend to derail an otherwise succesful
inflationary scenario by giving rise to scalar (mass)$^2$ terms of order
$H^2$, where $H$ denotes the Hubble constant. Remarkably, it turns out that for
a canonical SUGRA potential (with minimal K\"ahler potential
$|S|^2+|\Phi|^2+|\overline{\Phi}|^2$), the problematic (mass)$^2$ term cancels
out for the superpotential $W_1$ in Eq. (\ref{super}) \cite{Copeland:1994vg}.
This property also persists when non-renormalizable terms that are permitted by
the $U(1)_R$ symmetry are included in the superpotential.\footnote{In general,
$K$ is expanded as
$K=|S|^2+|\Phi|^2+|\overline{\Phi}|^2-\alpha|S|^4/m^2_P+\ldots$, and only the
$|S|^4$ term in $K$ generates a (mass)$^2$ for $S$, which would spoil
inflation for $\alpha\sim1$ \cite{Panagiotakopoulos:1997ej}.  From the requirement
$\sigma<m_P$, one obtains an upper bound on $\alpha$ ($\sim10^{-2}$
for $\kappa\gtrsim10^{-2}$, $\sim10^{-5}$ for $\kappa\sim10^{-5}$) \cite{Asaka:1999jb}.
We should note that, since the superpotential is linear in the inflaton, 
the presence of other fields with Planck scale VEVs
also induce an inflaton mass of order $H$. Some ways to suppress the inflaton mass
are discussed in \cite{Panagiotakopoulos:2004tf}.}

The scalar potential is given by
\begin{equation} \label{SUGRA}
V={\rm e}^K\left[\left(\frac{\partial^2 K}{\partial z_i\partial z^*_j}\right)^{-1}D_{z_i}W D_{z^*_j}W^*-3|W|^2\right]+V_D\,,
\end{equation}
with
\begin{equation} \label{SUGRA2}
D_{z_i}W=\frac{\partial W}{\partial z_i}+\frac{\partial K}{\partial z_i}W\,,
\end{equation}
where the sum extends over all fields $z_i$, and $K=\sum_i |z_i|^2$
is the minimal K\"ahler potential.  The D-term $V_D$ vanishes in the D-flat direction
$|\overline{\Phi}|=|\Phi|$. From Eq. (\ref{SUGRA}), with a minimal K\"ahler potential one contribution
to the inflationary potential is given by \cite{Copeland:1994vg,Panagiotakopoulos:1997qd,Linde:1997sj,Kawasaki:2003zv}
\begin{equation}
V_{\rm SUGRA}=\kappa^{2}M^{4}\left[\frac{|S|^{4}}{2}+\ldots\right]\,.
\end{equation}
There are additional contributions to the potential arising from the soft
SUSY breaking terms. In $N=1$ SUGRA these include the universal scalar masses
equal to $m_{3/2}$ ($\sim$ TeV), the gravitino mass.
However, their effect on the
inflationary scenario is negligible, as discussed below.
The more important term is the $A$ term
 $(2-A)m_{3/2}\kappa M^2 S(+\rm{h.c.})$. For convenience, we write this as
$a\,m_{3/2} \kappa M^2 |S|$, where $a\equiv2|2-A|\cos(\arg S+\arg(2-A))$.
The effective potential is approximately given by
Eq. (\ref{loop}) plus the leading SUGRA correction $\kappa^2 M^4 |S|^4/2$ and
the $A$ term:
{\setlength\arraycolsep{2pt}
\begin{eqnarray} \label{potential} V_{1}&=&\kappa^{2}M^{4}\bigg[1
+\frac{\kappa^{2}\mathcal{N}}{32\pi^{2}} \Big(
2\ln\frac{\kappa^{2}|S|^{2}}{\Lambda^{2}}+(z+1)^{2}\ln(1+z^{-1})\nonumber\\
& +&(z-1)^{2}\ln(1-z^{-1})\Big)+\frac{|S|^4}{2} \bigg]+ a\,m_{3/2} \kappa M^2|S|\,. \end{eqnarray}}
We perform our numerical calculations using this potential, taking $|a\,m_{3/2}|$=1 TeV.
It is, however, instructive to discuss small and large $\kappa$ limits of Eq. (\ref{potential}).
For $\kappa\gg10^{-3}$, $1\gg\sigma\gg \sqrt{2}M$, and Eq. (\ref{potential}) becomes
\begin{equation} \label{v1}
V_1\simeq\kappa^{2}M^{4}\left[1+\frac{\kappa^{2}\mathcal{N}}{32\pi^{2}}2\ln\frac{\kappa^{2}\sigma^{2}}{2\Lambda^{2}}+
\frac{\sigma^{4}}{8}\right] \end{equation}
to a good approximation. Comparing the derivatives of the radiative and SUGRA corrections
one sees that the radiative term dominates for
$\sigma^2\lesssim\kappa\sqrt{\mathcal{N}}/2\pi$. From $3H\dot{\sigma}=-V'$,
$\sigma^2_0\simeq\kappa^2 \mathcal{N}N_0/4\pi^2$ for the one-loop effective
potential, so that SUGRA effects are negligible only for
$\kappa\ll2\pi/\sqrt{\mathcal{N}}N_{0}\simeq 0.1/\sqrt{\mathcal{N}}$. (For
$\mathcal{N}=1$, this essentially agrees with \cite{Linde:1997sj}).

$\mathcal{P^{{\rm 1/2}}_R}$ is found from Eq. (\ref{v1}) to be
\begin{equation} \mathcal{P^{{\rm 1/2}}_R}\simeq\frac{1}{\sqrt{3}\pi}\frac{\kappa\,M^2}{\sigma^3_0}\,.  \end{equation}
\noindent In the absence of the SUGRA correction, the gauge symmetry breaking
scale $M$ is given by Eq. (\ref{quad}). For $\kappa\gg10^{-3}$, $x_0\gg1$ and
$x_{0}\,y_{0}\,f(x^{2}_{0})\to1^{-}$.  $\mathcal{P^{{\rm 1/2}}_R}$ in this case 
turns out to be proportional to $(M/m_{P})^2$\cite{Dvali:1994ms,Lazarides:2001zd}.  
Using the WMAP best fit $\mathcal{P^{{\rm 1/2}}_R}\simeq4.7\times10^{-5}$ \cite{Spergel:2003cb}, 
$M$ approaches the value $\mathcal{N}^{1/4}\cdot6\times10^{15}$ GeV. The presence of
the SUGRA term leads to larger values of $\sigma_0$ and hence larger values of
$M$ for $\kappa\gtrsim0.06/\sqrt{\mathcal{N}}$.

For $\kappa\ll10^{-3}$, $|S_0|\simeq M$ where $S_0$ is the value
of the field at $k_0$, i.e. $z\simeq1$.  (Note that due to the flatness of the
potential the last 55 or so e-folds occur with $|S|$ close to $M$.)
From Eqs. (\ref{perturb}, \ref{potential}), as $z\to1$
\begin{equation} \label{spert}
\mathcal{P^{{\rm 1/2}}_R}=\frac{2\sqrt{2}\pi}{\sqrt{3}}\frac{\kappa^2 M^4}{\ln(2)\kappa^3 M \mathcal{N}+8\pi^2 \kappa M^5
+4\pi^2 a\,m_{3/2}}\,.
\end{equation}
The denominator of Eq. (\ref{spert}) contains the radiative, SUGRA and the $A$
terms respectively.  Comparing them, we see that the radiative term can be
ignored for $\kappa\lesssim10^{-4}$. There is also a soft mass term $m^2_{3/2} |S|^2$ in the potential, 
corresponding to an additional term $8\pi^2 m^2_{3/2} /\kappa M$ in the denominator.
We have omitted this term, since it is insignificant
for $\kappa\gtrsim10^{-5}$.

For a positive $A$ term ($a>0$), the maximum value of $\mathcal{P^{{\rm 1/2}}_R}$
as a function of $M$ is found to be
\begin{equation} \label{rmax}
\mathcal{P^{{\rm 1/2}}_R}_{\rm{max}}=\frac{1}{2^{7/10}\, 5^{3/2}\, 3\pi}\left(\frac{\kappa^6}{a\,m_{3/2}}\right)^{1/5}\,.
\end{equation}
Setting $\mathcal{P^{{\rm
1/2}}_R}\simeq4.7\times10^{-5}$, we find a
lower bound on $\kappa$ ($\simeq10^{-5}$). For larger values of
$\kappa$, there are two separate solutions of $M$ for a given
$\kappa$. The solution with larger $M$ is not valid if the symmetry breaking pattern produces cosmic strings. 
For example, strings are produced when $\Phi,\,\overline{\Phi}$ break $U(1)_R\times U(1)_{B-L}$ to $U(1)_Y\times Z_2$ 
matter parity, but not when $\Phi,\,\overline{\Phi}$ are $SU(2)_R\times U(1)_{B-L}$ doublets. 
For $a<0$, there are again two solutions, but for the solution with
a lower value of $M$, the slope changes sign as the inflaton rolls
for $\kappa\lesssim10^{-4}$ and the inflaton gets trapped in a false
vacuum.

Note that the $A$ term depends on $\arg S$, so it
should be checked whether $\arg S$ changes significantly during inflation.
Numerically, we find that it does not, except for a range of $\kappa$ around
$10^{-4}$ \cite{Senoguz:2004vu}. For this range,
if the initial value of the $S$ field is greater than $M$ by at least a factor of two or so,
the $A$ term and the slope become negative even if they were initially positive, before inflation can
suitably end. However, larger values of the $A$ term, or the mass term coming from a non-minimal K\"ahler potential
(or from a hidden sector VEV) would drive the value of $M$ in that region up, allowing the slope to stay positive
(see \cite{Jeannerot:2005mc} for the effect of varying the $A$ term and the mass).

\begin{figure}[t]
\psfrag{k}{\footnotesize{$\kappa$}}
\includegraphics[height=.2\textheight]{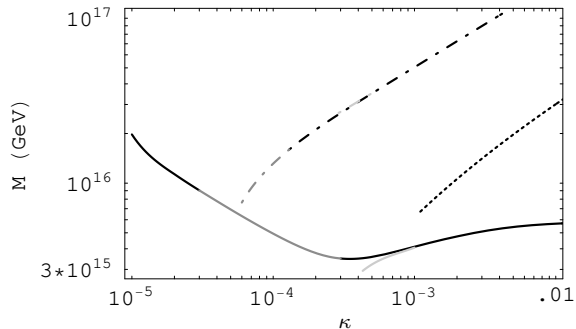}
\vspace{-.6cm}
\caption{  The value of the symmetry breaking scale $M$ vs.
$\kappa$, for SUSY hybrid inflation with $\mathcal{N}=1$ (solid),
and for shifted hybrid inflation
(dot-dashed for $M_S=m_P$, dotted for $M_S=5\times10^{17}$ GeV).
Light grey portions of the curves are for $a<0$, where only the
segments that do not overlap with the solutions for $a>0$ are shown.
The grey segments denote the range of $\kappa$ for
which the change in $\arg S$ is significant.} \label{fig:21}
\end{figure}

The dependence of $M$ on $\kappa$ is shown in Fig. \ref{fig:21}. 
Note that with inflation linked to the breaking of MSSM$\times U(1)_{B-L}$, $M$ 
corresponds to the $U(1)_{B-L}$ breaking scale, which
is not fixed by the evolution of the three SM gauge couplings.
The amplitude of the curvature perturbation (or, equivalently, $\delta T/T$)
determines this scale to be close to the SUSY GUT scale, suggesting that
$U(1)_{B-L}$ could be embedded in $SO(10)$ or its subgroups. For example,
$M$ can be determined in flipped $SU(5)$
                from the renormalization group evolution of the $SU(3)$ and
                $SU(2)$ gauge couplings. The values are remarkably consistent
                with the ones fixed from $\delta T/T$ considerations \cite{Kyae:2005nv}.

Here, some remarks concerning the allowed range of $\kappa$ is in order.
As discussed above, a lower bound on $\kappa$ is obtained from the inflationary dynamics
and the amplitude of the curvature perturbation.
An upper bound on $\kappa$ is obtained from the value of the spectral index, which
we discuss next. The gravitino constraint provides a more stringent upper bound ($\kappa\lesssim10^{-2}$), as discussed
in the next section. If cosmic strings form (as would be the case for $\mathcal{N}=1$), 
the range of $\kappa$ is also restricted by the limits on the cosmic string contribution 
to $\mathcal{P^{{\rm 1/2}}_R}$, however most of the range may still be allowed \cite{Jeannerot:2005mc}.

\begin{figure}[t]
\psfrag{k}{\footnotesize{$\kappa$}}
\includegraphics[height=.2\textheight]{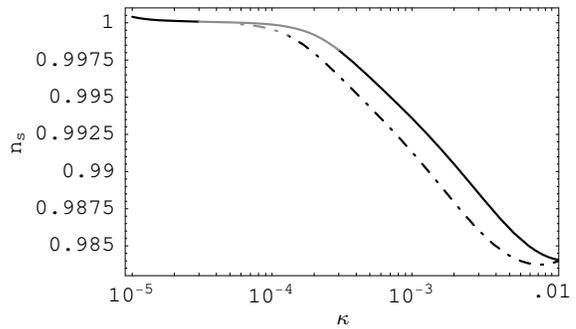}
\vspace{-.7cm}
\caption{  The spectral index $n_s$ vs. $\kappa$, for SUSY hybrid
inflation with $\mathcal{N}=1$ (solid), and for shifted hybrid inflation with $M_S=m_P$
(dot-dashed). The grey segments denote the range of $\kappa$ for
which the change in $\arg S$ is significant.} \label{fig:22}
\end{figure}

\begin{figure}[t]
\psfrag{k}{\footnotesize{$\kappa$}}
\includegraphics[height=.2\textheight]{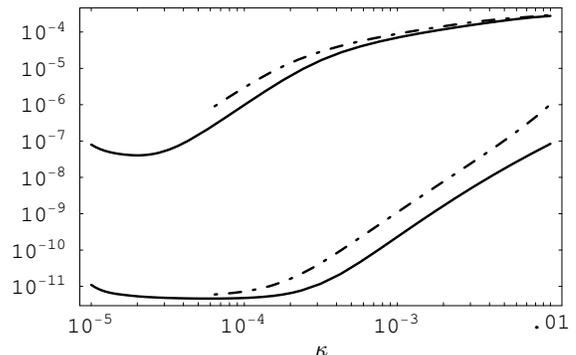}
\vspace{-.5cm}
\caption{$\textrm{d}n_{s}/\textrm{d}\ln k$ vs. $\kappa$ (top) and tensor to scalar ratio $r$ vs. $\kappa$ (bottom), 
for SUSY hybrid inflation with $\mathcal{N}=1$ (solid), and for shifted hybrid inflation with $M_S=m_P$
(dot-dashed).} \label{fig:rdn}
\end{figure}

In the absence of SUGRA corrections, the scalar spectral index $n_s$
for $\kappa\gg10^{-3}$ is given by \cite{Dvali:1994ms}
\begin{equation} n_{s}\simeq1+2\eta\simeq 1-\frac{1}{N_{0}}\simeq0.98\,,
\end{equation}
while it approaches unity for small $\kappa$.
When the SUGRA correction is taken into account,
from Eq. (\ref{v1}),
\begin{equation}
n_s\simeq1+2\eta\simeq1+2\left(3\sigma^2-\frac{\kappa^2\mathcal{N}}{8\pi^2\sigma^2}\right)\,,
\end{equation}
\noindent and it exceeds unity for
$\sigma^2\gtrsim\kappa\sqrt{\mathcal{N}}/2\sqrt{3}\pi$. For $x_0\gg1$,
\begin{equation} \label{nq}
N_0=\int_{\sigma_{end}}^{\sigma_0}\frac{V}{V'}\textrm{d}\sigma
\approx\frac{\pi}{2\sigma^2_0}\frac{\kappa}{\kappa_c}\tan\left(
\frac{\pi}{2}\frac{\kappa}{\kappa_c}\right)\,, \end{equation}
\noindent where
$\kappa_c=\pi^2/\sqrt{\mathcal{N}}N_0\simeq0.16/\sqrt{\mathcal{N}}$.  Using
Eq. (\ref{nq}), one finds that the spectral index $n_s$ exceeds unity for
$\kappa\simeq2\pi/\sqrt{3\mathcal{N}}N_0\simeq0.06/\sqrt{\mathcal{N}}$.  The
dependence of $n_s$ on $\kappa$ is displayed in Fig. \ref{fig:22}.
$\textrm{d}n_{s}/\textrm{d}\ln k$ is small and the tensor to scalar ratio $r$ is negligible,
as shown in Fig. \ref{fig:rdn}.

The experimental data
disfavor $n_s$ values in excess of unity on smaller scales (say
$k\lesssim0.05$ Mpc$^{-1}$), which leads to 
$\kappa\lesssim0.1/\sqrt{\mathcal{N}}$ for $n_s\le1.04$.\footnote{Larger values of $\kappa$
are allowed in models where dissipative effects are significant. Such effects
become important for large values of $\kappa$, provided the inflaton also
has strong couplings to matter fields \cite{Bastero-Gil:2004tg}.}
Thus, even for the largest allowed $\kappa$ the vacuum energy density during inflation is considerably smaller than
the symmetry breaking scale, and the tensor to scalar
ratio $r\lesssim10^{-4}$.

Note that the initial WMAP analysis suggests a running spectral index, with 
$|{\rm d}n_s/{\rm d}\ln k|\lesssim10^{-3}$ disfavored at
the $2\sigma$ level \cite{Spergel:2003cb}.
On the other hand, an analysis including the constraints
from the Sloan Digital Sky Survey (SDSS) finds no evidence for running, with
$n_s=0.98\pm0.02$ and ${\rm d}n_s/{\rm d}\ln k=-0.003\pm0.010$ \cite{Seljak:2004xh}.
 Clearly, more data is necessary to resolve this
important issue. Modifications of the models discussed here, generally involving two stages of inflation, 
have been proposed in \cite{Kawasaki:2003zv,Senoguz:2004ky} and elsewhere to generate a much more significant
variation of $n_s$ with $k$.

The inflationary scenario based on the superpotential $W_1$ in
Eq. (\ref{super}) has the characteristic feature that the end of
inflation essentially coincides with the gauge symmetry breaking.  Thus,
modifications should be made to $W_1$ if the breaking of $G$ to $H$ leads to
the appearance of topological defects such as monopoles or domain
walls. For instance, the breaking of $G_{PS}\equiv SU(4)_c\times SU(2)_L\times
SU(2)_R$ \cite{Pati:1974yy} to the MSSM by fields belonging to
$\Phi(\overline{4},1,2)$, $\overline{\Phi}(4,1,2)$ produces magnetic monopoles
that carry two quanta of Dirac magnetic charge \cite{Lazarides:1980cc}. As
shown in \cite{Jeannerot:2000sv}, one simple resolution of the topological defects problem
is achieved by supplementing $W_1$ with a non-renormalizable term:
\begin{equation} \label{super2} W_2=\kappa
S(\overline{\Phi}\Phi-v^{2})-\frac{S(\overline{\Phi}\Phi)^{2}}{M^{2}_{S}}\,,
\end{equation}
\noindent where $v$ is comparable to the SUSY GUT scale
$M_{\rm GUT}\simeq2\times10^{16}$ GeV and $M_{S}$ is an effective
cutoff scale. The dimensionless coefficient of the non-renormalizable term is
absorbed in $M_S$.  The presence of the non-renormalizable term enables an
inflationary trajectory along which the gauge symmetry is broken.  Thus, in
this `shifted' hybrid inflation model the topological defects are inflated away.

The inflationary potential is similar to Eq. (\ref{potential}) \cite{Jeannerot:2000sv}:
{\setlength\arraycolsep{2pt}
\begin{eqnarray} \label{potential2} V_{2}&=&
\kappa^{2}m^{4}\bigg[1+\frac{\kappa^{2}}{16\pi^{2}} \Big(
2\ln\frac{\kappa^{2}|S|^{2}}{\Lambda^{2}}+(z+1)^{2}\ln(1+z^{-1})\nonumber\\
&+&(z-1)^{2}\ln(1-z^{-1})\Big) +\frac{|S|^4}{2}\bigg]+a\,m_{3/2}\kappa v^2|S|{}\,.  \end{eqnarray}}
Here $m^{2}=v^{2}(1/4\xi-1)$ where $\xi=v^{2}/\kappa M^{2}_{S}$, $z\equiv x^{2}\equiv\sigma^{2}/m^{2}$,
and $2-A$ is replaced by $2-A+A/2\xi$ in the expression for $a$.\footnote{Note 
that the potential also contains a mass term even for minimal K\"ahler potential, 
due to the nonvanishing VEVs of $\Phi,\,\overline{\Phi}$.
This term, however, does not have a significant effect for the
range of parameters where $v\lesssim10^{17}$ GeV.}
Just like in the previous section, for $\kappa\lesssim0.01$ the SUGRA correction is negligible, and one
obtains Eq. (\ref{quad}) with $\mathcal{N}$ replaced by 2, Eq. (\ref{m_kap})
with $\mathcal{N}$ replaced by 4, and $M$ replaced by $m$ in both equations.
The slow roll parameters (and therefore $n_s$, ${\rm d}n_s/{\rm d}\ln k$, and
$r$) are identical to the SUSY hybrid inflation model with $\mathcal{N}=2$
for $z\gg1$ (Fig. \ref{fig:22}).

The VEV
$M=\big|\langle\nu^c_H\rangle\big|=\big|\langle\overline{\nu}^c_H\rangle\big|$ at
the SUSY minimum is given by \cite{Jeannerot:2000sv}
\begin{equation}
\left(\frac{M}{v}\right)^{2}=\frac{1}{2\xi}\left(1-\sqrt{1-4\xi}\right)\,,
\end{equation}
\noindent and is $\sim10^{16}-10^{17}$ GeV depending on $\kappa$ and $M_S$.
The system follows the inflationary trajectory for
$1/7.2<\xi<1/4$ \cite{Jeannerot:2000sv}, which is satisfied for
$\kappa\gtrsim10^{-5}$ if the effective cutoff scale $M_S=m_P$.
For lower values of $M_S$, the inflationary trajectory is followed only for
higher values of $\kappa$, and $M$ is lower for a given $\kappa$ (Fig. \ref{fig:21}).

A variation on these inflationary scenarios is obtained by imposing a
$Z_{2}$ symmetry on the superpotential, so that only even powers of the
combination $\Phi\overline{\Phi}$ are allowed \cite{Lazarides:1995vr}:
\begin{equation} \label{super3} W_3=S\left(-v^2
+\frac{(\Phi\overline{\Phi})^{2}}{M^{2}_{S}}\right)\,, \end{equation}
\noindent where the dimensionless parameter $\kappa$ is absorbed in $v$.  The resulting scalar
potential possesses two (symmetric) valleys of local minima which are suitable
for inflation and along which the GUT symmetry is broken. The
inclination of these valleys is already non-zero at the classical level and the
end of inflation is smooth, in contrast to inflation based on the
superpotential $W_1$ (Eq. (\ref{super})). An important consequence is that, as
in the case of shifted hybrid inflation, potential problems associated
with topological defects are avoided.

\begin{figure}[t] \includegraphics[height=.2\textheight]{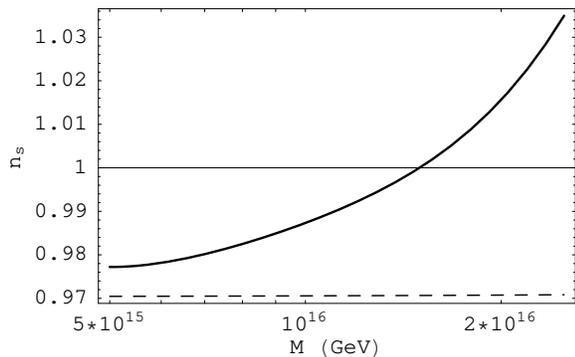} \vspace{-.5cm} \caption{The spectral index $n_s$
as a function of the gauge symmetry breaking
scale $M$ for smooth hybrid inflation (dashed line--without SUGRA
correction, solid line--with SUGRA correction).} \label{fig:24} 
\end{figure}

The common VEV at the SUSY minimum
$M=\big|\langle\nu^c_H\rangle\big|=\big|\langle\overline{\nu}^c_H\rangle\big|=
(v\,M_S)^{1/2}$. For $\sigma^{2}\gg M^2$, the inflationary potential
is given by
\begin{equation} \label{v3}
V_3\approx v^{4}\left[1-\frac{2}{27}\frac{M^4}{\sigma^{4}}+\frac{\sigma^4}{8}\right]\,,
\end{equation}
\noindent where the last term arises from the canonical SUGRA correction. The
soft terms in this case do not have a significant effect on the inflationary dynamics. If we
set $M$ equal to the SUSY GUT scale $M_{\rm GUT}$, we get $v\simeq1.4\times10^{15}$
GeV and $M_S\simeq2.8\times10^{17}$ GeV. (Note that, if we express
Eq. (\ref{super3}) in terms of the coupling parameter $\kappa$,
this value corresponds to $\kappa\sim O(v^2/M^2_{\rm GUT})\sim 10^{-2}$.) 
The value of the field $\sigma$ is $1.1\times10^{17}$ GeV at
the end of inflation (corresponding to $\eta=-1$) and is
$\sigma_0\simeq2.4\times10^{17}$ GeV at $k_0$.  In the absence of the SUGRA
correction (which is small for $M\lesssim10^{16}$ GeV), $\sigma\propto
M^{2/3}\,m_P^{1/3}$, $(\delta T/T)_0\propto M^{10/3}/(M^2_S\,m_P^{4/3})$ and
the spectral index is given by \cite{Lazarides:1995vr}
\begin{equation} n_{s}\simeq1-\frac{5}{3N_{0}}\simeq0.97\,, \end{equation}
\noindent a value which coincides with the prediction of some D-brane
inflation models \cite{Dvali:2001fw}. This may not be surprising since, in the
absence of SUGRA correction, the potential $V_3$ (Eq. (\ref{v3})) has a form
familiar from D-brane inflation.  The SUGRA correction raises $n_{s}$
from 0.97 to 1.0 for $M\sim10^{16}$ GeV, and above unity for
$M\gtrsim1.5\times10^{16}$ GeV (Fig. \ref{fig:24}).

\section{Leptogenesis In Supersymmetric Hybrid Inflation Models}
\label{chap3}
An important constraint on SUSY hybrid inflation models
arises from considering the reheat temperature $T_{r}$
after inflation, taking into account the gravitino problem which requires that
$T_{r}\lesssim10^6$--$10^{11}$ GeV \cite{Khlopov:1984pf}. This
constraint on $T_r$ depends on the SUSY breaking mechanism and the
gravitino mass $m_{3/2}$. For gravity mediated SUSY breaking models
with unstable gravitinos of mass $m_{3/2}\simeq0.1$--1 TeV,
$T_r\lesssim10^6$--$10^9$ GeV \cite{Kawasaki:1995af}, while
$T_r\lesssim10^{10}$ GeV for stable gravitinos \cite{Bolz:2000fu}.  In gauge
mediated models the reheat temperature is generally more severely constrained, although
$T_r\sim10^9$--$10^{10}$ GeV is possible for $m_{3/2}\simeq5$--100 GeV
\cite{Gherghetta:1998tq}. Finally, the anomaly mediated symmetry breaking (AMSB)
scenario may allow gravitino masses much heavier than a TeV, thus
accommodating a reheat temperature as high as $10^{11}$ GeV \cite{Gherghetta:1999sw}.

After the end of inflation in the models discussed in section~\ref{chap2}, the
fields fall toward the SUSY vacuum and perform damped oscillations about it.
The vevs of $\overline{\Phi}$, $\Phi$ along their right handed neutrino
components $\overline{\nu}^c_H$, $\nu^c_H$  break the gauge symmetry.  The
oscillating system, which we collectively denote as $\chi$, consists of the two
complex scalar fields $(\delta\overline{\nu}^c_H+\delta\nu^c_H)/\sqrt{2}$
(where $\delta\overline{\nu}^c_H$, $\delta\nu^c_H$ are the deviations of
$\overline{\nu}^c_H$, $\nu^c_H$ from $M$) and $S$, with equal mass $m_{\chi}$.

We assume here that the inflaton $\chi$ decays predominantly into right handed neutrino
superfields $N_i$, via the superpotential coupling $(1/m_P)\gamma_{ij}
\overline{\phi}\,\overline{\phi}N_i N_j$ or $\gamma_{ij}\overline{\phi} N_i N_j$,
where $i,j$ are family indices 
(see below for a different scenario connected to the resolution of the MSSM $\mu$
problem). Their subsequent out of equilibrium decay to lepton and Higgs
superfields generates lepton asymmetry, which is then partially converted into
the observed baryon asymmetry by sphaleron effects
\cite{Fukugita:1986hr}. 

The right handed neutrinos, as shown below, can be heavy compared to
the reheat temperature $T_r$.
Note that unlike thermal leptogenesis, there is then no washout factor
since lepton number violating 2-body scatterings
mediated by right handed neutrinos are out of equilibrium as long as the
lightest right handed neutrino mass $M_1\gg T_r$ \cite{Fukugita:1990gb}.  More
precisely, the washout factor is proportional to $e^{-z}$ where $z=M_1/T_r$ \cite{Buchmuller:2003gz},
and can be neglected for $z\gtrsim10$.
Without this assumption, the constraints to generate sufficient lepton
asymmetry would be more stringent.

GUTs typically relate the Dirac neutrino masses to that of the quarks or charged leptons.
It is therefore reasonable to assume the Dirac masses are hierarchical. The
low-energy neutrino data indicates that the right handed neutrinos in this case
will also be hierarchical in general.  As discussed in Ref.
\cite{Akhmedov:2003dg}, setting the Dirac masses strictly equal to the up-type
quark masses and fitting to the neutrino oscillation parameters generally
yields strongly hierarchical right handed neutrino masses ($M_1\ll M_2\ll
M_3$), with $M_1\sim10^5$ GeV. 
The lepton asymmetry in this case is too small
by several orders of magnitude. However, it is plausible that there are large
radiative corrections to the first two family Dirac masses, so that $M_1$ remains 
heavy compared to $T_r$.

\begin{figure}[t] 
\psfrag{m}{\scriptsize{$m_{\chi}$ (GeV)}}
\psfrag{k}{\footnotesize{$\kappa$}}
\includegraphics[height=.2\textheight]{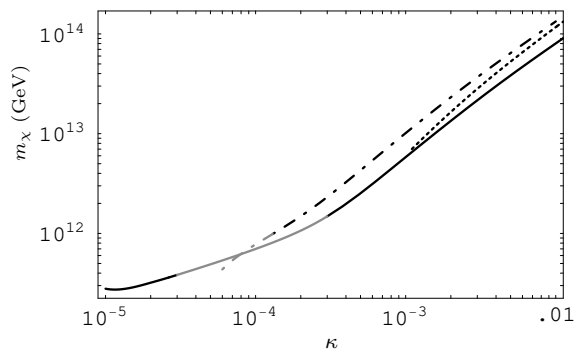} 
\vspace{-0.5cm} 
\caption{\sf The inflaton mass $m_{\chi}$ vs. $\kappa$,  
for SUSY hybrid inflation with $\mathcal{N}=1$ (solid), 
and for shifted hybrid inflation 
(dot-dashed for $M_S=m_P$, dotted for $M_S=5\times10^{17}$ GeV). 
The grey segments denote the range of $\kappa$ for which the change in $\arg S$ is significant.} \label{minf}
\end{figure}

\begin{figure}[ht] 
\psfrag{m}{\scriptsize{$m_{\chi}$ (GeV)}}
\includegraphics[height=.2\textheight]{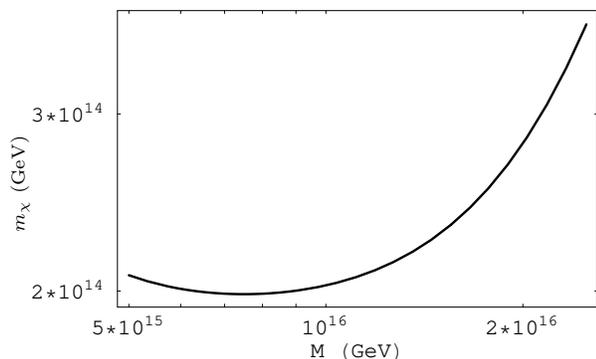} 
 \vspace{-0.5cm} 
\caption{\sf The inflaton mass $m_{\chi}$ vs. the symmetry breaking scale $M$ for smooth hybrid inflation.} \label{minf2}
\end{figure}

\begin{figure}[t] 
\psfrag{k}{\footnotesize{$\kappa$}}
\includegraphics[height=.2\textheight]{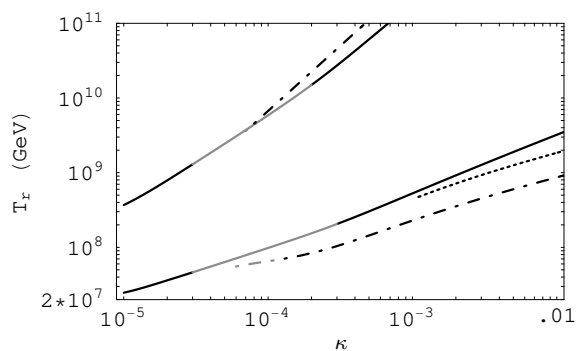} 
\vspace{-0.5cm} 
\caption{\sf The lower bound on the reheat temperature $T_r$ vs. $\kappa$,  
for SUSY hybrid inflation with $\mathcal{N}=1$ (solid) and for shifted hybrid inflation 
(dot-dashed for $M_S=m_P$, dotted for $M_S=5\times10^{17}$ GeV). The segments in the top left part of the figure correspond to the bounds in the presence of a $\lambda S h^2$ coupling. The grey segments denote the range of $\kappa$ for which the change in $\arg S$ is significant.} \label{ktr}
\end{figure}

\begin{figure}[htb] 
\includegraphics[height=.2\textheight]{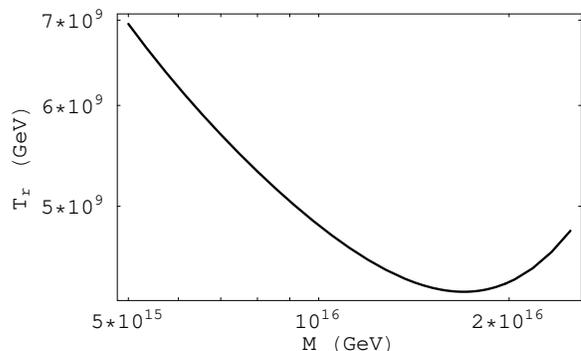}
\vspace{-0.5cm} 
\caption{\sf The lower bound on the reheat temperature $T_r$ vs. the symmetry breaking scale $M$ for smooth hybrid inflation.} \label{sm_mtr}
\end{figure}

A reasonable mass pattern is therefore $M_1<M_2\ll M_3$, which can result from either
the dimensionless couplings $\gamma_{ij}$ or additional symmetries (see e.g. \cite{Senoguz:2004ky}).
The dominant contribution to the lepton asymmetry is
still from the decays with $N_3$ in the loop, as long as the first two family
right handed neutrinos are not quasi degenerate.
Under these assumptions, the lepton asymmetry is given by \cite{Asaka:1999yd}
\begin{equation} \label{nls}
\frac{n_L}{s}\lesssim3\times10^{-10}\frac{T_r}{m_{\chi}}\left(\frac{M_i}{10^6\rm{\
GeV}}\right)\left(\frac{m_{\nu3}}{0.05\rm{\ eV}}\right)\,, \end{equation}
where $M_i$ denotes the mass of the heaviest right handed neutrino the inflaton
can decay into. 
The decay rate $\Gamma_{\chi}=(1/8\pi)(M^2_i/M^2)m_{\chi}$
\cite{Lazarides:1997dv}, and the reheat temperature $T_r$ is given by
\begin{equation} \label{reheat} T_r=\left(\frac{45}{2\pi^2
g_*}\right)^{1/4}(\Gamma_\chi\,m_P)^{1/2}\simeq
0.063\frac{(m_P\,m_{\chi})^{1/2}}{M}M_i \,.\end{equation}
From the experimental value of the
baryon to photon ratio $\eta_B\simeq6.1\times10^{-10}$ \cite{Spergel:2003cb},
the required lepton asymmetry is found to be $n_L/s\simeq2.5\times10^{-10}$
\cite{Khlebnikov:1988sr}.
Using this value, along with Eqs. (\ref{nls}, \ref{reheat}), we can express $T_r$ in terms of
the symmetry breaking scale $M$ and the inflaton mass $m_{\chi}$:
\begin{eqnarray} \label{trmin}
T_r&\gtrsim&\left(\frac{10^{16}{\rm\
GeV}}{M}\right)^{1/2}\left(\frac{m_{\chi}}{10^{11}{\rm\ GeV}}\right)^{3/4}\nonumber\\
&\times&\left(\frac{0.05\rm{\ eV}}{m_{\nu3}}\right)^{1/2}
1.6\times10^{7}{\rm\ GeV}
\,.  \end{eqnarray}
Here $m_{\chi}$ is given by $\sqrt{2}\kappa M$, $\sqrt{2}\kappa M \sqrt{1-4\xi}$ and
$2\sqrt{2}v^2/M$ respectively for hybrid, shifted hybrid and smooth hybrid inflation.
The value of $m_{\chi}$ is shown in Figs. \ref{minf} and \ref{minf2}.
We show the lower bound on $T_r$ calculated using this equation (taking $m_{\nu3}=0.05$ eV) 
in Figs. \ref{ktr}, \ref{sm_mtr}. 

Eq. (\ref{reheat}) also yields the result that the heaviest right handed
neutrino the inflaton can decay into is about 400 (6) times heavier than
$T_r$, for hybrid inflation with $\kappa=10^{-5}$ ($10^{-2}$). For shifted hybrid inflation, this ratio
does not depend on $\kappa$ as strongly and
is $\sim10^2$ \cite{Senoguz:2003hc}. This is consistent with ignoring washout
effects as long as the lightest right handed neutrino mass $M_1$ is also $\gg T_r$.

Both the gravitino constraint and the constraint $M_1\gg T_r$ favor smaller
values of $\kappa$ for hybrid inflation, with
$T_r\gtrsim2\times10^7$ GeV for $\kappa\sim10^{-5}$. 
Similarly, the gravitino constraint favors $\kappa$ values as small as the inflationary trajectory
allows for shifted hybrid inflation, and $T_r\gtrsim5\times10^7$ GeV for $M_S=m_P$.
Smooth hybrid inflation is
relatively disfavored since $T_r\gtrsim4\times10^9$ and $M_2/T_r\simeq4\ (12)$
for $M=5\times10^{15}$ GeV ($2\times10^{16}$ GeV).\footnote{A new inflation model related to
smooth hybrid inflation is discussed in \cite{Asaka:1999jb,Senoguz:2004ky}, where
the energy scale of inflation is lower and consequently lower
reheat temperatures are allowed.}

So far we have not addressed the $\mu$ problem and the relationship to $T_r$ in the present context.
The MSSM $\mu$ problem can naturally be resolved in SUSY hybrid inflation models
in the presence of the term $\lambda Sh^2$ in the superpotential, where
$h$ contains the two Higgs doublets \cite{Dvali:1998uq}. (The `bare' term $h^2$ is not allowed
by the $U(1)$ R-symmetry.) After inflation
the VEV of $S$ generates a $\mu$ term with $\mu=\lambda\langle S\rangle=-m_{3/2}\lambda/\kappa$,
where $\lambda>\kappa$ is required for the proper vacuum.
The inflaton in this case predominantly decays into higgses (and higgsinos) with $\Gamma_h=(1/16\pi)
\lambda^2 m_{\chi}$. As a consequence the presence of this term
significantly increases the reheat temperature $T_r$. Following \cite{Lazarides:1998qx},
we calculate $T_r$ for the best case scenario
$\lambda=\kappa$. We find  a lower bound on $T_r$ of $4\times10^{8}$ GeV in hybrid inflation,
see Fig. \ref{ktr}. $T_r\gtrsim4\times10^9$ GeV for shifted hybrid inflation with
$M_S=m_P$\,\footnote{We take $\lambda/\kappa=2/(1/4\xi-1)$ for shifted hybrid inflation. 
Some scalars belonging to the inflaton sector acquire negative (mass)$^2$
if $\lambda$ is smaller. $\kappa\sim10^{-4}$ corresponds to $\lambda/\kappa\simeq3$.}, 
and $T_r\gtrsim10^{12}$ GeV for smooth hybrid inflation.
An alternative resolution of the $\mu$ problem in SUSY hybrid inflation involves
a Peccei-Quinn (PQ) symmetry $U(1)_{\rm PQ}$ \cite{Lazarides:1998iq,Jeannerot:2000sv}.

The lower bounds on $T_r$ are summarized in Table 1. 
Even for hybrid and
shifted models, there is some tension between the gravitino constraint and the
reheat temperature required to generate sufficient lepton asymmetry,
particularly for gravity mediated SUSY breaking models, and if hadronic decays
of gravitinos are not suppressed.
However, we should note that 
having quasi degenerate neutrinos would increase the lepton asymmetry per neutrino
decay $\epsilon$ \cite{Flanz:1996fb} and thus allow lower values of $T_r$ corresponding to lighter right
handed neutrinos. Provided that the neutrino mass splittings are comparable to
their decay widths, $\epsilon$ can be as large as $1/2$ \cite{Pilaftsis:1997jf}.
The lepton asymmetry in this case is of order $T_r/m_{\chi}$ where $m_{\chi}\sim10^{11}$
GeV for $\kappa\sim10^{-5}$, and sufficient lepton asymmetry
can be generated with $T_r$ close to the electroweak scale.

\begin{table}[t]
\caption{Lower bounds on the reheat temperature (GeV)}
\begin{tabular}{l@{\hspace{1cm}}r@{\hspace{1cm}}r}
\toprule
 & without $\lambda S h^2$ & with $\lambda S h^2$ \\
\colrule
SUSY hybrid inflation & $2\times10^7$ & $4\times10^8$ \\
\colrule
Shifted hybrid inflation & $5\times10^7$ & $4\times10^9$ \\
\colrule
Smooth hybrid inflation & $4\times10^9$ & $\gtrsim10^{12}$ \\
\botrule
\end{tabular}
\label{ttable}
\end{table}

\section{Summary}
\begin{itemize}
\item MSSM$\,\times Z_2$ (matter parity), when extended to MSSM$\,\times U(1)_{B-L}\times U(1)_R$
    helps us to implement inflation and leptogenesis; natural extensions
    include $SO(10)$ and its various subgroups;

\item The $U(1)_{B-L}$ breaking scale is constrained from $\delta T/T$ to be of order
the SUSY GUT scale;

\item The spectral index is predicted to be $0.99\pm0.01$, with a negligible tensor to scalar ratio
and $\textrm {d}n_{s}/\textrm {d}\ln k\lesssim10^{-3}$. For smooth hybrid inflation $n_s\ge0.97$;

\item $U(1)_R$ contains $Z_2$ matter parity, eliminates Planck suppressed dimension
    five proton decay, and may play an essential role in resolving the MSSM $\mu$
    problem.

\end{itemize}
\section{Acknowledgements}
This work is partially supported by the US Department of Energy under contract number DE-FG02-91ER40626.


\begin{thebibliography}{52}
\expandafter\ifx\csname natexlab\endcsname\relax\def\natexlab#1{#1}\fi
\expandafter\ifx\csname bibnamefont\endcsname\relax
  \def\bibnamefont#1{#1}\fi
\expandafter\ifx\csname bibfnamefont\endcsname\relax
  \def\bibfnamefont#1{#1}\fi
\expandafter\ifx\csname citenamefont\endcsname\relax
  \def\citenamefont#1{#1}\fi
\expandafter\ifx\csname url\endcsname\relax
  \def\url#1{\texttt{#1}}\fi
\expandafter\ifx\csname urlprefix\endcsname\relax\def\urlprefix{URL }\fi
\providecommand{\bibinfo}[2]{#2}
\providecommand{\eprint}[2][]{\url{#2}}

\bibitem[{\citenamefont{{\c S}eno$\breve{\textrm{g}}$uz and
  Shafi}(2003)}]{Senoguz:2003zw}
\bibinfo{author}{\bibfnamefont{V.~N.} \bibnamefont{{\c
  S}eno$\breve{\textrm{g}}$uz}} \bibnamefont{and}
  \bibinfo{author}{\bibfnamefont{Q.}~\bibnamefont{Shafi}},
  \bibinfo{journal}{Phys. Lett.} \textbf{\bibinfo{volume}{B567}},
  \bibinfo{pages}{79} (\bibinfo{year}{2003}), \eprint{hep-ph/0305089}.

\bibitem[{\citenamefont{{\c S}eno$\breve{\textrm{g}}$uz and
  Shafi}(2005)}]{Senoguz:2004vu}
\bibinfo{author}{\bibfnamefont{V.~N.} \bibnamefont{{\c
  S}eno$\breve{\textrm{g}}$uz}} \bibnamefont{and}
  \bibinfo{author}{\bibfnamefont{Q.}~\bibnamefont{Shafi}},
  \bibinfo{journal}{Phys. Rev.} \textbf{\bibinfo{volume}{D71}},
  \bibinfo{pages}{043514} (\bibinfo{year}{2005}), \eprint{hep-ph/0412102}.

\bibitem[{\citenamefont{Lazarides}(2002)}]{Lazarides:2001zd}
\bibinfo{author}{\bibfnamefont{G.}~\bibnamefont{Lazarides}},
  \bibinfo{journal}{Lect. Notes Phys.} \textbf{\bibinfo{volume}{592}},
  \bibinfo{pages}{351} (\bibinfo{year}{2002}), \eprint{hep-ph/0111328}.

\bibitem[{\citenamefont{Dvali et~al.}(1994)\citenamefont{Dvali, Shafi, and
  Schaefer}}]{Dvali:1994ms}
\bibinfo{author}{\bibfnamefont{G.~R.} \bibnamefont{Dvali}},
  \bibinfo{author}{\bibfnamefont{Q.}~\bibnamefont{Shafi}}, \bibnamefont{and}
  \bibinfo{author}{\bibfnamefont{R.~K.} \bibnamefont{Schaefer}},
  \bibinfo{journal}{Phys. Rev. Lett.} \textbf{\bibinfo{volume}{73}},
  \bibinfo{pages}{1886} (\bibinfo{year}{1994}), \eprint{hep-ph/9406319}.

\bibitem[{\citenamefont{Copeland et~al.}(1994)\citenamefont{Copeland, Liddle,
  Lyth, Stewart, and Wands}}]{Copeland:1994vg}
\bibinfo{author}{\bibfnamefont{E.~J.} \bibnamefont{Copeland}},
  \bibinfo{author}{\bibfnamefont{A.~R.} \bibnamefont{Liddle}},
  \bibinfo{author}{\bibfnamefont{D.~H.} \bibnamefont{Lyth}},
  \bibinfo{author}{\bibfnamefont{E.~D.} \bibnamefont{Stewart}},
  \bibnamefont{and} \bibinfo{author}{\bibfnamefont{D.}~\bibnamefont{Wands}},
  \bibinfo{journal}{Phys. Rev.} \textbf{\bibinfo{volume}{D49}},
  \bibinfo{pages}{6410} (\bibinfo{year}{1994}), \eprint{astro-ph/9401011}.

\bibitem[{\citenamefont{Linde}(1991)}]{Linde:1991km}
\bibinfo{author}{\bibfnamefont{A.~D.} \bibnamefont{Linde}},
  \bibinfo{journal}{Phys. Lett.} \textbf{\bibinfo{volume}{B259}},
  \bibinfo{pages}{38} (\bibinfo{year}{1991}).

\bibitem[{\citenamefont{Liddle and Lyth}(1993)}]{Liddle:1993fq}
\bibinfo{author}{\bibfnamefont{A.~R.} \bibnamefont{Liddle}} \bibnamefont{and}
  \bibinfo{author}{\bibfnamefont{D.~H.} \bibnamefont{Lyth}},
  \bibinfo{journal}{Phys. Rept.} \textbf{\bibinfo{volume}{231}},
  \bibinfo{pages}{1} (\bibinfo{year}{1993}), \eprint{astro-ph/9303019}.

\bibitem[{\citenamefont{Liddle and Lyth}(1992)}]{Liddle:1992wi}
\bibinfo{author}{\bibfnamefont{A.~R.} \bibnamefont{Liddle}} \bibnamefont{and}
  \bibinfo{author}{\bibfnamefont{D.~H.} \bibnamefont{Lyth}},
  \bibinfo{journal}{Phys. Lett.} \textbf{\bibinfo{volume}{B291}},
  \bibinfo{pages}{391} (\bibinfo{year}{1992}), \eprint{astro-ph/9208007}.

\bibitem[{\citenamefont{Lazarides et~al.}(1997)\citenamefont{Lazarides,
  Schaefer, and Shafi}}]{Lazarides:1997dv}
\bibinfo{author}{\bibfnamefont{G.}~\bibnamefont{Lazarides}},
  \bibinfo{author}{\bibfnamefont{R.~K.} \bibnamefont{Schaefer}},
  \bibnamefont{and} \bibinfo{author}{\bibfnamefont{Q.}~\bibnamefont{Shafi}},
  \bibinfo{journal}{Phys. Rev.} \textbf{\bibinfo{volume}{D56}},
  \bibinfo{pages}{1324} (\bibinfo{year}{1997}), \eprint{hep-ph/9608256}.

\bibitem[{\citenamefont{Panagiotakopoulos}(1997{\natexlab{a}})}]{Panagiotakopo%
ulos:1997ej}
\bibinfo{author}{\bibfnamefont{C.}~\bibnamefont{Panagiotakopoulos}},
  \bibinfo{journal}{Phys. Lett.} \textbf{\bibinfo{volume}{B402}},
  \bibinfo{pages}{257} (\bibinfo{year}{1997}{\natexlab{a}}),
  \eprint{hep-ph/9703443}.

\bibitem[{\citenamefont{Asaka et~al.}(2000)\citenamefont{Asaka, Hamaguchi,
  Kawasaki, and Yanagida}}]{Asaka:1999jb}
\bibinfo{author}{\bibfnamefont{T.}~\bibnamefont{Asaka}},
  \bibinfo{author}{\bibfnamefont{K.}~\bibnamefont{Hamaguchi}},
  \bibinfo{author}{\bibfnamefont{M.}~\bibnamefont{Kawasaki}}, \bibnamefont{and}
  \bibinfo{author}{\bibfnamefont{T.}~\bibnamefont{Yanagida}},
  \bibinfo{journal}{Phys. Rev.} \textbf{\bibinfo{volume}{D61}},
  \bibinfo{pages}{083512} (\bibinfo{year}{2000}), \eprint{hep-ph/9907559}.

\bibitem{Panagiotakopoulos:2004tf}
  C.~Panagiotakopoulos,
  Phys.\ Rev.\  {\bf D71}, 063516 (2005),
  hep-ph/0411143.

\bibitem[{\citenamefont{Panagiotakopoulos}(1997{\natexlab{b}})}]{Panagiotakopo%
ulos:1997qd}
\bibinfo{author}{\bibfnamefont{C.}~\bibnamefont{Panagiotakopoulos}},
  \bibinfo{journal}{Phys. Rev.} \textbf{\bibinfo{volume}{D55}},
  \bibinfo{pages}{7335} (\bibinfo{year}{1997}{\natexlab{b}}),
  \eprint{hep-ph/9702433}.

\bibitem[{\citenamefont{Linde and Riotto}(1997)}]{Linde:1997sj}
\bibinfo{author}{\bibfnamefont{A.~D.} \bibnamefont{Linde}} \bibnamefont{and}
  \bibinfo{author}{\bibfnamefont{A.}~\bibnamefont{Riotto}},
  \bibinfo{journal}{Phys. Rev.} \textbf{\bibinfo{volume}{D56}},
  \bibinfo{pages}{1841} (\bibinfo{year}{1997}), \eprint{hep-ph/9703209}.

\bibitem[{\citenamefont{Kawasaki et~al.}(2003)\citenamefont{Kawasaki,
  Yamaguchi, and Yokoyama}}]{Kawasaki:2003zv}
\bibinfo{author}{\bibfnamefont{M.}~\bibnamefont{Kawasaki}},
  \bibinfo{author}{\bibfnamefont{M.}~\bibnamefont{Yamaguchi}},
  \bibnamefont{and} \bibinfo{author}{\bibfnamefont{J.}~\bibnamefont{Yokoyama}},
  \bibinfo{journal}{Phys. Rev.} \textbf{\bibinfo{volume}{D68}},
  \bibinfo{pages}{023508} (\bibinfo{year}{2003}), \eprint{hep-ph/0304161}.

\bibitem[{\citenamefont{Spergel et~al.}(2003)}]{Spergel:2003cb}
\bibinfo{author}{\bibfnamefont{D.~N.} \bibnamefont{Spergel}}
  \bibnamefont{et~al.}, \bibinfo{journal}{Astrophys. J. Suppl.}
  \textbf{\bibinfo{volume}{148}}, \bibinfo{pages}{175} (\bibinfo{year}{2003}),
  \eprint{astro-ph/0302209};
\bibinfo{author}{\bibfnamefont{H.~V.} \bibnamefont{Peiris}}
  \bibnamefont{et~al.}, \bibinfo{journal}{Astrophys. J. Suppl.}
  \textbf{\bibinfo{volume}{148}}, \bibinfo{pages}{213} (\bibinfo{year}{2003}),
  \eprint{astro-ph/0302225}.

\bibitem{Jeannerot:2005mc}
  R.~Jeannerot and M.~Postma,
  JHEP {\bf 0505}, 071 (2005),
  hep-ph/0503146.

\bibitem{Kyae:2005nv}
  B.~Kyae and Q.~Shafi,
  hep-ph/0510105.

\bibitem{Bastero-Gil:2004tg}
  M.~Bastero-Gil and A.~Berera,
  Phys.\ Rev.\  {\bf D71}, 063515 (2005),
  hep-ph/0411144.

\bibitem{Seljak:2004xh}
  U.~Seljak {\it et al.},
  Phys.\ Rev.\  {\bf D71}, 103515 (2005),
  astro-ph/0407372.

\bibitem[{\citenamefont{{\c S}eno$\breve{\textrm{g}}$uz and
  Shafi}(2004{\natexlab{a}})}]{Senoguz:2004ky}
\bibinfo{author}{\bibfnamefont{V.~N.} \bibnamefont{{\c
  S}eno$\breve{\textrm{g}}$uz}} \bibnamefont{and}
  \bibinfo{author}{\bibfnamefont{Q.}~\bibnamefont{Shafi}},
  \bibinfo{journal}{Phys. Lett.} \textbf{\bibinfo{volume}{B596}},
  \bibinfo{pages}{8} (\bibinfo{year}{2004}{\natexlab{a}}),
  \eprint{hep-ph/0403294}.

\bibitem[{\citenamefont{Pati and Salam}(1974)}]{Pati:1974yy}
\bibinfo{author}{\bibfnamefont{J.~C.} \bibnamefont{Pati}} \bibnamefont{and}
  \bibinfo{author}{\bibfnamefont{A.}~\bibnamefont{Salam}},
  \bibinfo{journal}{Phys. Rev.} \textbf{\bibinfo{volume}{D10}},
  \bibinfo{pages}{275} (\bibinfo{year}{1974}).

\bibitem[{\citenamefont{Lazarides et~al.}(1980)\citenamefont{Lazarides, Magg,
  and Shafi}}]{Lazarides:1980cc}
\bibinfo{author}{\bibfnamefont{G.}~\bibnamefont{Lazarides}},
  \bibinfo{author}{\bibfnamefont{M.}~\bibnamefont{Magg}}, \bibnamefont{and}
  \bibinfo{author}{\bibfnamefont{Q.}~\bibnamefont{Shafi}},
  \bibinfo{journal}{Phys. Lett.} \textbf{\bibinfo{volume}{B97}},
  \bibinfo{pages}{87} (\bibinfo{year}{1980}).

\bibitem[{\citenamefont{Jeannerot et~al.}(2000)\citenamefont{Jeannerot, Khalil,
  Lazarides, and Shafi}}]{Jeannerot:2000sv}
\bibinfo{author}{\bibfnamefont{R.}~\bibnamefont{Jeannerot}},
  \bibinfo{author}{\bibfnamefont{S.}~\bibnamefont{Khalil}},
  \bibinfo{author}{\bibfnamefont{G.}~\bibnamefont{Lazarides}},
  \bibnamefont{and} \bibinfo{author}{\bibfnamefont{Q.}~\bibnamefont{Shafi}},
  \bibinfo{journal}{JHEP} \textbf{\bibinfo{volume}{10}}, \bibinfo{pages}{012}
  (\bibinfo{year}{2000}), \eprint{hep-ph/0002151}.

\bibitem[{\citenamefont{Lazarides and
  Panagiotakopoulos}(1995)}]{Lazarides:1995vr}
\bibinfo{author}{\bibfnamefont{G.}~\bibnamefont{Lazarides}} \bibnamefont{and}
  \bibinfo{author}{\bibfnamefont{C.}~\bibnamefont{Panagiotakopoulos}},
  \bibinfo{journal}{Phys. Rev.} \textbf{\bibinfo{volume}{D52}},
  \bibinfo{pages}{559} (\bibinfo{year}{1995}), \eprint{hep-ph/9506325};
%
\bibinfo{author}{\bibfnamefont{G.}~\bibnamefont{Lazarides}},
  \bibinfo{author}{\bibfnamefont{C.}~\bibnamefont{Panagiotakopoulos}},
  \bibnamefont{and} \bibinfo{author}{\bibfnamefont{N.~D.}
  \bibnamefont{Vlachos}}, \bibinfo{journal}{Phys. Rev.}
  \textbf{\bibinfo{volume}{D54}}, \bibinfo{pages}{1369} (\bibinfo{year}{1996}),
  \eprint{hep-ph/9606297};
%
\bibinfo{author}{\bibfnamefont{R.}~\bibnamefont{Jeannerot}},
  \bibinfo{author}{\bibfnamefont{S.}~\bibnamefont{Khalil}}, \bibnamefont{and}
  \bibinfo{author}{\bibfnamefont{G.}~\bibnamefont{Lazarides}},
  \bibinfo{journal}{Phys. Lett.} \textbf{\bibinfo{volume}{B506}},
  \bibinfo{pages}{344} (\bibinfo{year}{2001}), \eprint{hep-ph/0103229}.

\bibitem[{\citenamefont{Dvali et~al.}(2001)\citenamefont{Dvali, Shafi, and
  Solganik}}]{Dvali:2001fw}
\bibinfo{author}{\bibfnamefont{G.~R.} \bibnamefont{Dvali}},
  \bibinfo{author}{\bibfnamefont{Q.}~\bibnamefont{Shafi}}, \bibnamefont{and}
  \bibinfo{author}{\bibfnamefont{S.}~\bibnamefont{Solganik}}
  (\bibinfo{year}{2001}), \eprint{hep-th/0105203};
%
\bibinfo{author}{\bibfnamefont{C.~P.} \bibnamefont{Burgess}}
  \bibnamefont{et~al.}, \bibinfo{journal}{JHEP} \textbf{\bibinfo{volume}{07}},
  \bibinfo{pages}{047} (\bibinfo{year}{2001}), \eprint{hep-th/0105204};
%
\bibinfo{author}{\bibfnamefont{B.} \bibnamefont{Kyae}} \bibnamefont{and}
  \bibinfo{author}{\bibfnamefont{Q.}~\bibnamefont{Shafi}},
  \bibinfo{journal}{Phys. Lett.} \textbf{\bibinfo{volume}{B526}},
  \bibinfo{pages}{379} (\bibinfo{year}{2002}), \eprint{hep-ph/0111101};
%
\bibinfo{author}{\bibfnamefont{C.}~\bibnamefont{Herdeiro}},
  \bibinfo{author}{\bibfnamefont{S.}~\bibnamefont{Hirano}}, \bibnamefont{and}
  \bibinfo{author}{\bibfnamefont{R.}~\bibnamefont{Kallosh}},
  \bibinfo{journal}{JHEP} \textbf{\bibinfo{volume}{12}}, \bibinfo{pages}{027}
  (\bibinfo{year}{2001}), \eprint{hep-th/0110271};
%
  J.~H.~Brodie and D.~A.~Easson,
  JCAP {\bf 0312}, 004 (2003),
  hep-th/0301138.

\bibitem[{\citenamefont{Khlopov and Linde}(1984)}]{Khlopov:1984pf}
\bibinfo{author}{\bibfnamefont{M.~Y.} \bibnamefont{Khlopov}} \bibnamefont{and}
  \bibinfo{author}{\bibfnamefont{A.~D.} \bibnamefont{Linde}},
  \bibinfo{journal}{Phys. Lett.} \textbf{\bibinfo{volume}{B138}},
  \bibinfo{pages}{265} (\bibinfo{year}{1984}).

\bibitem[{\citenamefont{Kawasaki and Moroi}(1995)}]{Kawasaki:1995af}
\bibinfo{author}{\bibfnamefont{M.}~\bibnamefont{Kawasaki}} \bibnamefont{and}
  \bibinfo{author}{\bibfnamefont{T.}~\bibnamefont{Moroi}},
  \bibinfo{journal}{Prog. Theor. Phys.} \textbf{\bibinfo{volume}{93}},
  \bibinfo{pages}{879} (\bibinfo{year}{1995}), \eprint{hep-ph/9403364}.

\bibitem[{\citenamefont{Bolz et~al.}(2001)\citenamefont{Bolz, Brandenburg, and
  Buchmuller}}]{Bolz:2000fu}
\bibinfo{author}{\bibfnamefont{M.}~\bibnamefont{Bolz}},
  \bibinfo{author}{\bibfnamefont{A.}~\bibnamefont{Brandenburg}},
  \bibnamefont{and}
  \bibinfo{author}{\bibfnamefont{W.}~\bibnamefont{Buchmuller}},
  \bibinfo{journal}{Nucl. Phys.} \textbf{\bibinfo{volume}{B606}},
  \bibinfo{pages}{518} (\bibinfo{year}{2001}), \eprint{hep-ph/0012052};
%
\bibinfo{author}{\bibfnamefont{M.}~\bibnamefont{Fujii}},
  \bibinfo{author}{\bibfnamefont{M.}~\bibnamefont{Ibe}}, \bibnamefont{and}
  \bibinfo{author}{\bibfnamefont{T.}~\bibnamefont{Yanagida}},
  \bibinfo{journal}{Phys. Lett.} \textbf{\bibinfo{volume}{B579}},
  \bibinfo{pages}{6} (\bibinfo{year}{2004}), \eprint{hep-ph/0310142};
%
\bibinfo{author}{\bibfnamefont{L.}~\bibnamefont{Roszkowski}},
  \bibinfo{author}{\bibfnamefont{R.}~\bibnamefont{Ruiz~de Austri}},
  \bibnamefont{and} \bibinfo{author}{\bibfnamefont{K.-Y.} \bibnamefont{Choi}},
  \bibinfo{journal}{JHEP} \textbf{\bibinfo{volume}{08}}, \bibinfo{pages}{080}
  (\bibinfo{year}{2005}), \eprint{hep-ph/0408227}.

\bibitem[{\citenamefont{Gherghetta
  et~al.}(1999{\natexlab{a}})\citenamefont{Gherghetta, Giudice, and
  Riotto}}]{Gherghetta:1998tq}
\bibinfo{author}{\bibfnamefont{T.}~\bibnamefont{Gherghetta}},
  \bibinfo{author}{\bibfnamefont{G.~F.} \bibnamefont{Giudice}},
  \bibnamefont{and} \bibinfo{author}{\bibfnamefont{A.}~\bibnamefont{Riotto}},
  \bibinfo{journal}{Phys. Lett.} \textbf{\bibinfo{volume}{B446}},
  \bibinfo{pages}{28} (\bibinfo{year}{1999}{\natexlab{a}}),
  \eprint{hep-ph/9808401}.

\bibitem[{\citenamefont{Gherghetta
  et~al.}(1999{\natexlab{b}})\citenamefont{Gherghetta, Giudice, and
  Wells}}]{Gherghetta:1999sw}
\bibinfo{author}{\bibfnamefont{T.}~\bibnamefont{Gherghetta}},
  \bibinfo{author}{\bibfnamefont{G.~F.} \bibnamefont{Giudice}},
  \bibnamefont{and} \bibinfo{author}{\bibfnamefont{J.~D.} \bibnamefont{Wells}},
  \bibinfo{journal}{Nucl. Phys.} \textbf{\bibinfo{volume}{B559}},
  \bibinfo{pages}{27} (\bibinfo{year}{1999}{\natexlab{b}}),
  \eprint{hep-ph/9904378}.

\bibitem[{\citenamefont{Fukugita and Yanagida}(1986)}]{Fukugita:1986hr}
\bibinfo{author}{\bibfnamefont{M.}~\bibnamefont{Fukugita}} \bibnamefont{and}
  \bibinfo{author}{\bibfnamefont{T.}~\bibnamefont{Yanagida}},
  \bibinfo{journal}{Phys. Lett.} \textbf{\bibinfo{volume}{B174}},
  \bibinfo{pages}{45} (\bibinfo{year}{1986}). For non-thermal
leptogenesis, see 
\bibinfo{author}{\bibfnamefont{G.}~\bibnamefont{Lazarides}} \bibnamefont{and}
  \bibinfo{author}{\bibfnamefont{Q.}~\bibnamefont{Shafi}},
  \bibinfo{journal}{Phys. Lett.} \textbf{\bibinfo{volume}{B258}},
  \bibinfo{pages}{305} (\bibinfo{year}{1991}{\natexlab{a}}).


\bibitem[{\citenamefont{Fukugita and Yanagida}(1990)}]{Fukugita:1990gb}
\bibinfo{author}{\bibfnamefont{M.}~\bibnamefont{Fukugita}} \bibnamefont{and}
  \bibinfo{author}{\bibfnamefont{T.}~\bibnamefont{Yanagida}},
  \bibinfo{journal}{Phys. Rev.} \textbf{\bibinfo{volume}{D42}},
  \bibinfo{pages}{1285} (\bibinfo{year}{1990}).

\bibitem[{\citenamefont{Buchmuller et~al.}(2003)\citenamefont{Buchmuller,
  Di~Bari, and Plumacher}}]{Buchmuller:2003gz}
\bibinfo{author}{\bibfnamefont{W.}~\bibnamefont{Buchmuller}},
  \bibinfo{author}{\bibfnamefont{P.}~\bibnamefont{Di~Bari}}, \bibnamefont{and}
  \bibinfo{author}{\bibfnamefont{M.}~\bibnamefont{Plumacher}},
  \bibinfo{journal}{Nucl. Phys.} \textbf{\bibinfo{volume}{B665}},
  \bibinfo{pages}{445} (\bibinfo{year}{2003}), \eprint{hep-ph/0302092}.

\bibitem[{\citenamefont{Akhmedov et~al.}(2003)\citenamefont{Akhmedov, Frigerio,
  and Smirnov}}]{Akhmedov:2003dg}
\bibinfo{author}{\bibfnamefont{E.~K.} \bibnamefont{Akhmedov}},
  \bibinfo{author}{\bibfnamefont{M.}~\bibnamefont{Frigerio}}, \bibnamefont{and}
  \bibinfo{author}{\bibfnamefont{A.~Y.} \bibnamefont{Smirnov}},
  \bibinfo{journal}{JHEP} \textbf{\bibinfo{volume}{09}}, \bibinfo{pages}{021}
  (\bibinfo{year}{2003}), \eprint{hep-ph/0305322}.

\bibitem[{\citenamefont{Asaka et~al.}(1999)\citenamefont{Asaka, Hamaguchi,
  Kawasaki, and Yanagida}}]{Asaka:1999yd}
\bibinfo{author}{\bibfnamefont{T.}~\bibnamefont{Asaka}},
  \bibinfo{author}{\bibfnamefont{K.}~\bibnamefont{Hamaguchi}},
  \bibinfo{author}{\bibfnamefont{M.}~\bibnamefont{Kawasaki}}, \bibnamefont{and}
  \bibinfo{author}{\bibfnamefont{T.}~\bibnamefont{Yanagida}},
  \bibinfo{journal}{Phys. Lett.} \textbf{\bibinfo{volume}{B464}},
  \bibinfo{pages}{12} (\bibinfo{year}{1999}), \eprint{hep-ph/9906366}.

\bibitem[{\citenamefont{Khlebnikov and Shaposhnikov}(1988)}]{Khlebnikov:1988sr}
\bibinfo{author}{\bibfnamefont{S.~Y.} \bibnamefont{Khlebnikov}}
  \bibnamefont{and} \bibinfo{author}{\bibfnamefont{M.~E.}
  \bibnamefont{Shaposhnikov}}, \bibinfo{journal}{Nucl. Phys.}
  \textbf{\bibinfo{volume}{B308}}, \bibinfo{pages}{885} (\bibinfo{year}{1988}).

\bibitem[{\citenamefont{{\c S}eno$\breve{\textrm{g}}$uz and
  Shafi}(2004{\natexlab{b}})}]{Senoguz:2003hc}
\bibinfo{author}{\bibfnamefont{V.~N.} \bibnamefont{{\c
  S}eno$\breve{\textrm{g}}$uz}} \bibnamefont{and}
  \bibinfo{author}{\bibfnamefont{Q.}~\bibnamefont{Shafi}},
  \bibinfo{journal}{Phys. Lett.} \textbf{\bibinfo{volume}{B582}},
  \bibinfo{pages}{6} (\bibinfo{year}{2004}{\natexlab{b}}),
  \eprint{hep-ph/0309134}.

\bibitem[{\citenamefont{Dvali et~al.}(1998)\citenamefont{Dvali, Lazarides, and
  Shafi}}]{Dvali:1998uq}
\bibinfo{author}{\bibfnamefont{G.~R.} \bibnamefont{Dvali}},
  \bibinfo{author}{\bibfnamefont{G.}~\bibnamefont{Lazarides}},
  \bibnamefont{and} \bibinfo{author}{\bibfnamefont{Q.}~\bibnamefont{Shafi}},
  \bibinfo{journal}{Phys. Lett.} \textbf{\bibinfo{volume}{B424}},
  \bibinfo{pages}{259} (\bibinfo{year}{1998}), \eprint{hep-ph/9710314};
%
\bibinfo{author}{\bibfnamefont{S.~F.} \bibnamefont{King}} \bibnamefont{and}
  \bibinfo{author}{\bibfnamefont{Q.}~\bibnamefont{Shafi}},
  \bibinfo{journal}{Phys. Lett.} \textbf{\bibinfo{volume}{B422}},
  \bibinfo{pages}{135} (\bibinfo{year}{1998}), \eprint{hep-ph/9711288}.

\bibitem[{\citenamefont{Lazarides and Vlachos}(1998)}]{Lazarides:1998qx}
\bibinfo{author}{\bibfnamefont{G.}~\bibnamefont{Lazarides}} \bibnamefont{and}
  \bibinfo{author}{\bibfnamefont{N.~D.} \bibnamefont{Vlachos}},
  \bibinfo{journal}{Phys. Lett.} \textbf{\bibinfo{volume}{B441}},
  \bibinfo{pages}{46} (\bibinfo{year}{1998}), \eprint{hep-ph/9807253}.

\bibitem[{\citenamefont{Lazarides and Shafi}(1998)}]{Lazarides:1998iq}
\bibinfo{author}{\bibfnamefont{G.}~\bibnamefont{Lazarides}} \bibnamefont{and}
  \bibinfo{author}{\bibfnamefont{Q.}~\bibnamefont{Shafi}},
  \bibinfo{journal}{Phys. Rev.} \textbf{\bibinfo{volume}{D58}},
  \bibinfo{pages}{071702} (\bibinfo{year}{1998}), \eprint{hep-ph/9803397}.

\bibitem[{\citenamefont{Flanz et~al.}(1996)\citenamefont{Flanz, Paschos,
  Sarkar, and Weiss}}]{Flanz:1996fb}
\bibinfo{author}{\bibfnamefont{M.}~\bibnamefont{Flanz}},
  \bibinfo{author}{\bibfnamefont{E.~A.} \bibnamefont{Paschos}},
  \bibinfo{author}{\bibfnamefont{U.}~\bibnamefont{Sarkar}}, \bibnamefont{and}
  \bibinfo{author}{\bibfnamefont{J.}~\bibnamefont{Weiss}},
  \bibinfo{journal}{Phys. Lett.} \textbf{\bibinfo{volume}{B389}},
  \bibinfo{pages}{693} (\bibinfo{year}{1996}), \eprint{hep-ph/9607310}. 

\bibitem[{\citenamefont{Pilaftsis}(1997)}]{Pilaftsis:1997jf}
\bibinfo{author}{\bibfnamefont{A.}~\bibnamefont{Pilaftsis}},
  \bibinfo{journal}{Phys. Rev.} \textbf{\bibinfo{volume}{D56}},
  \bibinfo{pages}{5431} (\bibinfo{year}{1997}), \eprint{hep-ph/9707235}.


\end{thebibliography}
\end{document}